\documentclass{ws-ijmpb}

\begin{document}

\markboth{F. Pan, G.-Y. Lu, and J. P. Draayer}
{Classification and Quantification
of Entangled Bipartite Qutrit Pure States}

%
\catchline{}{}{}{}{}
%

\title{Classification and Quantification
of Entangled \\Bipartite Qutrit Pure States}

\author{ Feng Pan$^{*}$, Guoying Lu}

\address{Department of Physics, Liaoning Normal University, Dalian
116029, P. R. China\\
$^{*}$daipan@dlut.edu.cn}

\author{J. P. Draayer}

\address{Department of Physics and Astronomy,
Louisiana State University, Baton Rouge, LA 70803-4001, USA\\
draayer@sura.org}

\maketitle

\begin{history}
\received{Day Month Year}
\revised{Day Month Year}
\end{history}

\begin{abstract}
A complete analysis of entangled bipartite qutrit pure states is carried
out based on a simple entanglement measure. An analysis of all possible
extremally entangled pure bipartite qutrit states is shown to reduce, with
the help of SLOCC transformations, to three distinct types. The analysis
and the results should be helpful for finding different entanglement types
in multipartite pure state systems.
\end{abstract}


Quantum entanglement is an important concept in quantum information
processing and quantum communication protocols.$^{1}$ Interest in
multi-dimensional entangled states comes from the foundations of quantum
mechanics as well as the development of new protocols in quantum
communication.  For example, it has been shown that maximally entangled
states of two quantum systems in a high-dimensional Hilbert space, qudits,
violate local realism stronger than qubits, and entangled qudits are less
affected by noise than entangled qubits.$^{2,3}$ In quantum
cryptography,$^{4}$ the use of entangled qutrits$^{4, 5}$ or qudits$^{6,7}$
instead of qubits is more secure against eavesdropping attacks.
Furthermore, the protocols for quantum teleportation or for quantum
cryptography work best with maximally entangled states. These facts
motivate the development of techniques to generate entangled states among
quantum systems in a higher dimensional Hilbert space with good
entanglement characteristics. Technical developments in this direction have
been made. For example, four polarized entangled photons have been used to
form two entangled qutrits.$^{8}$ Entangled qutrits with two photons using
an unbalanced 3-arm fiber optic interferometer or photonic orbital angular
momentum have been demonstrated.$^{9,10}$ Time-bin entangled qudits of up to
$11$ dimensions from pump pulses generated by a mode-locked laser have also
been reported.$^{11}$ In short, quantifying the entanglement measure of a
qutrit system is of physical interest.

In this paper, a complete analysis of entangled bipartite qutrit pure states
will be carried out based on a simple entanglement measure$^{12}$ and a
recent study of the classification of triqubit entangled states.$^{13}$
An analysis of all possible extremally entangled bipartite qutrit pure
states with up to six terms is shown to reduce, with the help of stocastic
local operation and classical communication (SLOCC) transformations, to
three distinct types. The analysis presented here, together with other
previously reported results, should be helpful in finding different
entanglement types in other multipartite qudit pure state systems.

Based on the method shown in Refs. 13 and 14, only entanglement properties
at the single copy level will be considered; asymptotic properties will not
be discussed. In such cases, it is well known that two pure states can
always be transformed with certainty from each other by means of a LOCC if
and only if they are related by a Local Unitary transformation (LU).
However, even in bipartite cases, entangled states are not always related
by a LU, and continuous parameters are needed to label all equivalence
classes. Hence, it seems that one needs to deal with infinitely many kinds
of entanglement. Fortunately, such arbitrariness has been overcome with the
help of SLOCC.$^{15}$ As defined in Ref. 15, states
$\Psi$ and $\Phi$ are equivalent under SLOCC if an Invertible Local
Operator (ILO) relating the two states exists, which is denoted as $Q_A$
and $Q_B$ for the ILOs for particles $A$ and $B$, respectively. Typically,
these ILOs are elements of the complex general linear group $GL_{A}(2,
c)\otimes GL_{B}(2, c)$ where each copy operates on the corresponding local
basis.

According to the definition,$^{12}$ for a $N$-particle
qutrit pure state $\psi$, the entanglement measure
can be defined as

$$\eta(\Psi)=
\left\{
\begin{array}{cc}
{1\over{N}}\sum^{N}_{i=1}
{S_{i}}&{{\rm if}~~S_i\neq0~\forall~i}\\
0&{\rm{if}}~~S_i=0,~~~~\end{array}
\right.\eqno(1)$$
where $S_i=-Tr[(\rho_{\Psi})_i \log_3{(\rho_{\Psi})_i}]$
is the reduced von Neumann entropy for the ith particle only,
with the other $N-1$ particles traced out,
and $(\rho_{\Psi})_i$ is the
corresponding reduced density matrix.
In the case of $N=2$, (1) reduces to
the well-known entanglement measure
for bipartite pure states, in which
we use the logarithm to the base $3$ instead of base $2$
to ensure that the maximal measure is normalized to $1$
in the qutrit case.
It can be verified
that the state $\Psi$ is partially separable when one of the reduced von
Neumann entropies $S_i$ is zero. In such cases the state $\Psi$ is not a
genuine entangled N-qutrit state. Furthermore, definition (1) is invariant
under LU, which is equivalent to LOCC for a pure state system.$^{12,13}$

Instead of qubit state, single-particle qutrit states can be denoted by
$|1\rangle$, $|0\rangle$, and $|-1\rangle$, which are assumed to be
mutually orthonormal. Product states of a bipartite qutrit system are
denoted by

$$\begin{array}{llll}
&&\{|U_1\rangle=\vert11\rangle,~~|U_2\rangle=\vert10\rangle,~~|U_3\rangle=\vert1-1\rangle,\\
&&|V_1\rangle=\vert01\rangle,~~|V_2\rangle=\vert00\rangle,~~|V_3\rangle=\vert0-1\rangle,\\
&&|W_1\rangle=\vert-11\rangle,~~|W_2\rangle=\vert-10\rangle,~~|W_3\rangle=\vert-1-1\rangle\}.
\end{array}\eqno(3)$$
These configurations span a $9$-dimensional Hilbert subspace and any
bipartite qutrit pure state can be expanded in terms of them.

\vskip .2cm
\hskip 0.3cm For two-term cases,
when a state is a linear combination of
a pair of states among \{($\vert U_i\rangle,~\vert
U_j\rangle$); ($\vert V_i\rangle,~\vert V_j\rangle$); ($\vert
W_i\rangle,~\vert W_j\rangle$)\}, or ($\vert U_i\rangle,~\vert
V_i\rangle$); ($\vert U_i\rangle,~\vert W_i\rangle$); ($\vert
V_i\rangle,~\vert W_i\rangle$), where $i,j \in
P$, $i\neq j$, $P=\{1,2,3\}$, it can always be decomposed
into a bipartite product state. Such states are separable
and disentangled. There are $18$ such combinations.
When a state is a linear combination of a pair of
states among ($\vert U_i\rangle,~\vert{V_j}\rangle$); ($\vert
U_i\rangle,~\vert{W_j}\rangle$); or ($\vert
V_i\rangle,~\vert{W_j}\rangle$), it is a genuinely entangled
state, where $i,j \in P$, $i\neq j$, $P=\{1,2,3\}$.
There are also $18$ such linear combinations in total.
Take $\vert\Psi\rangle
=a \vert U_1\rangle+b e^{i \alpha}\vert {V_2}\rangle= a
\vert11\rangle+b e^{i \alpha}\vert00\rangle$
as an example, where $a,~b\neq 0$ are real and satisfy
the normalization condition $a^2+b^2=1$, and
$\alpha$ is a relative phase.
Then, to maximize its measure (1) with the normalization
condition, one can find parameters for the corresponding extremal
cases. It can be verified easily that $\eta = 0.63093$
when $a=b={1\over\sqrt{2}}$ in such cases, and there is no
restriction on relative phase. We call such configurations Type I
entangled bipartite qutrit states.

\vskip .2cm \hskip 0.3cm For three term cases, if a state is a
linear combination of ($\vert U_i\rangle,~\vert U_j\rangle,~
\vert{U_k}\rangle$); ($\vert
V_i\rangle,~\vert{V_j}\rangle,~\vert{V_k}\rangle$);  ($\vert
W_i\rangle,~\vert{W_j}\rangle,~\vert{W_k}\rangle$), where $i,j,k\in
P$ and  $i\neq j\neq k$, or ($\vert U_i\rangle,~\vert V_i\rangle,~
\vert{W_i}\rangle$), where $i\in P$, $P=\{1,2,3\}$, it can always be
decomposed into a bipartite product state. Therefore, these
combinations are also separable. When a state is $a\vert
U_i\rangle+b e^{i \alpha}\vert {V_j}\rangle+c e^{i \beta}\vert
{W_k}\rangle$, where $a,~b,~c\neq 0$  are real and satisfy the
normalization condition $a^2+b^2+c^2=1$, and $\alpha,~\beta$ are
relative phases, it is called a Type III$_{1}$ state. There are $6$
such equivalent linear combinations. To maximize its measure (1)
with the constraint $a^2 + b^2 + c^2 = 1$, one can find parameters
for the corresponding extremal cases. It can be verified easily that
$\eta_{\max} =1$ when $a=b=c={1\over\sqrt{3}}$ for such cases, where
there is no restriction on relative phase. It can be proven that the
Type III$_{1}$ states are inequivalent to the Type I states. To do
so, we choose one representative of Type III$_{1}$ states,
$$\vert\Psi\rangle
=x_1\vert11\rangle+x_2\vert00\rangle+x_3\vert-1-1\rangle
=(x_1\vert11\rangle+x_2\vert00\rangle) +x_3\vert-1-1\rangle
\eqno(4)$$ with nonzero $x_1$, $x_2$, and $x_{3}$. It is clear that
the first two terms in (4) form a Type I state, which remains Type I
under the SLOCC, while the last term is always a bipartite product
state. Since all coefficients are nonzero in the case considered,
the state in (4) remains a linear combination of Type I entangled
state and a bipartite product state. It will degenerate into a Type
I state only when $x_3\rightarrow0$. Thus, we have proven that Type
III$_{1}$ is inequivalent to a Type I state, which will be called
Type II bipartite entangled qutrit states.

If a state is a linear combination of
($\vert
U_i\rangle, \vert{U_j}\rangle, \vert V_k\rangle$ or $\vert{
W_k}\rangle$); ($\vert V_i\rangle,~\vert{V_j}\rangle,~\vert
U_k\rangle$ or $\vert{ W_k}\rangle$); ($\vert
W_i\rangle,~\vert{W_j}\rangle,~\vert U_k\rangle$ or
$\vert{V_k}\rangle$),
where $i,~j\in P$, $k=i$ or $k=j$ and $i\neq j$, $P=\{1,2,3\}$,
it is called a Type III$_{2}$ state.
There are a total of $36$ such combinations.
In such cases, no extremal value of $\eta$ with all
coefficients non zero exists.
When a state is a linear combination of
  \{($\vert U_i\rangle,~\vert U_j\rangle,~\vert V_k\rangle$
or $\vert{W_k}\rangle$), ($\vert V_i\rangle,~\vert V_j\rangle,~\vert
U_k\rangle$ or $\vert{W_k}\rangle$), ($\vert W_i\rangle,~\vert
W_j\rangle,~\vert U_k\rangle$ or $\vert{V_k}\rangle$)\},
or \{($\vert U_i\rangle,\vert V_i\rangle,\vert W_j\rangle$),
($\vert{U_i}\rangle,\vert{W_i}\rangle,\vert{V_j}\rangle$)\},
($\vert{V_i}\rangle,\vert{W_i}\rangle,\vert{U_j}\rangle$)\}, where
$i,~j,~k\in P$,$i\neq j\neq k$,
$P=\{1,2,3\}$, it is called a Type III$_{3}$ state.
There are $36$ such combinations.
Take $\vert\Psi\rangle
=a \vert U_1\rangle+b e^{i \alpha}\vert {U_2}\rangle+c e^{i
\beta}\vert {V_3}\rangle= a \vert11\rangle+b e^{i
\alpha}\vert10\rangle+c e^{i \beta}\vert0-1\rangle $
as an example, where $a,~b,~c \neq 0$ are real
and satisfy the normalization condition $a^2+b^2+c^2=1$,
and $\alpha,~\beta$ are relative phases.
It can be shown with conditional maximization that
the extremal value of $\eta=0.63093$ in this case
when $a^2+b^2={1\over{2}}$ and $c={1\over{2}}$,
which is independent of the relative phases.
Using SLOCC, one can prove that
Type III$_{3}$ states are equivalent to
the Type I states.
\vskip .2cm
\hskip 0.3cm
For four term cases, there are $(^9_4)=126$ linear combinations
chosen from (3) that cannot be reduced into bipartite
product states. All these states are classified into $5$
types and listed in Table 1, for which there is no extremal
value of $\eta$ for the Type IV$_{1}$, IV$_{2}$, and IV$_{3}$
cases with all coefficients nonzero. Actually, these states
degenerate into Type I states under SLOCC.
For the Type IV$_{4}$ case, we take $\vert\Psi\rangle
=a \vert U_1\rangle+b e^{i \alpha}\vert {U_2}
\rangle+c e^{i \beta}\vert {V_1}\rangle+d e^{i \gamma}\vert {V_2}\rangle$
as an example, where $a,~b,~c,~d\neq 0$ are real and satisfy
the normalization condition $a^2+b^2+c^2+d^2=1$, and
$\alpha,~\beta,~\gamma$ are relative phases.
One can verify that $\eta=0.63093$ is its extremal value when
$a=d$, $b=c$, and $\varphi=\alpha+\beta-\gamma=2k\pi+\pi$.
Therefore, such states are also equivalent to
Type I entangled states under SLOCC.
For the Type IV$_{5}$ case, we take
$\vert\Psi\rangle
=a \vert U_1\rangle+b e^{i \alpha}
\vert {U_2}\rangle+c e^{i \beta}\vert {V_3}\rangle+d e^{i
\gamma}\vert {W_3}\rangle$
as an example, where $a,~b,~c,~d\neq 0$ are real and satisfy
the normalization condition $a^2+b^2+c^2+d^2=1$, and
$\alpha,~\beta,~\gamma$ are relative phases.
It can be shown that there is an extremal value
with $\eta=0.63093$ when  $a^2+b^2={1\over{2}}$ and
$c^2+d^2={1\over{2}}$, which is independent of the relative
phases. It can also be proven that these Type IV$_{5}$ states
can be transformed into Type I states under SLOCC.
Thus, four-term bipartite qutrit states are all equivalent
to the Type I states; no new type of entangled states are found
in this case.

For five term cases, there are also a total of $(^9_5)=126$ linear
combinations, of which all are genuine entangled states. They are
classified into $6$ types as shown in Table 2. For the Type V$_{1}$
case, we take $\vert\Psi\rangle =a \vert U_1\rangle+b e^{i
\alpha}\vert {U_2}\rangle+c e^{i \beta}\vert {U_3}\rangle +d e^{i
\gamma}\vert {V_1}\rangle+f e^{i \xi}\vert {V_2}\rangle$ as an
example, where $a,~b,~c,~d,~f\neq 0$ are real and satisfy
normalization condition $a^2+b^2+c^2+d^2+f^2=1$, and
$\alpha,~\beta,~\gamma,~\xi$ are relative phase factors. It can be
verified that $\eta=0.63093$ when $a^2+b^2+c^2={1\over2}$,
$d^2+f^2={1\over{2}}$, and $\varphi=\alpha+\gamma-\xi=2k\pi+\pi$.
Such a state is equivalent to a Type I entangled state under SLOCC.
For the Type V$_{2}$ case, we take $\vert\Psi\rangle =a \vert
U_1\rangle+b e^{i \alpha}\vert {U_2}\rangle+c e^{i \beta}\vert
{V_1}\rangle +d e^{i \gamma}\vert {V_2}\rangle+f e^{i \xi}\vert
{W_3}\rangle$ as an example, where $a,~b,~c,~d,~f\neq 0$ are real
and satisfy the normalization condition $a^2+b^2+c^2+d^2+f^2=1$, and
$\alpha,~\beta,~\gamma,~\xi$ are relative phases. It can be proven
that there is an extremal value $\eta=0.63093$ when
$a=b=c=d=\frac{1}{2\sqrt{2}}$, $f=\frac{1}{\sqrt{2}}$, and
$\varphi=\alpha+\beta-\gamma=2k\pi$. One can prove that Type V$_{2}$
states are equivalent to Type I entangled states under SLOCC. In
addition, there is another extremal value $\eta=1$

\begin{center}
\begin{center}{\bf Table 1.$~~~$Classification of entangled states
with four terms }\\
\end{center}\vskip .4cm
{\small
    \begin{small}
  \begin{tabular}{cccc}\hline\hline
  {Types}&~&Linear Combinations \\
  \hline
  \\
{IV$_{1}$}&$(|U_i\rangle,\vert{U_j}\rangle,\vert U_k\rangle,\vert V_m\rangle)$
&($|U_i\rangle,\vert{U_j}\rangle,\vert U_k\rangle,\vert W_m\rangle)$
&($|V_i\rangle,\vert{V_j}\rangle,\vert V_k\rangle,\vert U_m\rangle$)
\\\\
&($|V_i\rangle,\vert{V_j}\rangle,\vert V_k\rangle,\vert W_m\rangle$) &
($|W_i\rangle,\vert{W_j}\rangle,\vert W_k\rangle,\vert U_m\rangle$) &
($|W_i\rangle,\vert{W_j}\rangle,\vert W_k\rangle,\vert V_m\rangle$)\\
&  {$i,~j,~k,~m\in P$,}&{ $P=\{1,2,3\}$,}&{ $i\neq j\neq k$.} \\\\
&($|U_i\rangle,\vert{U_j}\rangle,\vert V_i\rangle,\vert W_i\rangle$)&
($|U_i\rangle,\vert{U_j}\rangle,\vert V_j\rangle,\vert W_j\rangle$)&
($|V_i\rangle,\vert{V_j}\rangle,\vert U_i\rangle,\vert W_i\rangle$)
\\\\
&($|V_i\rangle,\vert{V_j}\rangle,\vert U_j\rangle,\vert W_j\rangle$) &
($|W_i\rangle,\vert{W_j}\rangle,\vert U_i\rangle,\vert V_i\rangle$) &
($|W_i\rangle,\vert{W_j}\rangle,\vert U_j\rangle,\vert V_j\rangle$) \\
& {$i,~j\in P$,}&{$P=\{1,2,3\}$,}&{$i\neq j$.}\\\\
\hline\\
 {IV$_{2}$}&($|U_i\rangle,\vert{U_j}\rangle,\vert
V_i\rangle,\vert V_k\rangle$) & ($\vert U_i\rangle,\vert
U_j\rangle,\vert {V_j}\rangle,\vert {V_k}\rangle$)
&($\vert U_i\rangle,\vert U_j\rangle,\vert {W_i}\rangle,\vert
{W_k}\rangle$)\\\\
&($\vert U_i\rangle,\vert U_j\rangle,\vert {W_j}\rangle,\vert {W_k}\rangle$)
&($\vert V_i\rangle,\vert V_j\rangle,\vert {W_i}\rangle,\vert {W_k}\rangle$)
&($\vert V_i\rangle,\vert V_j\rangle,\vert {W_j}\rangle,\vert {W_k}\rangle$)\\
& {$i,~j,~k\in P$,}&{ $P=\{1,2,3\}$,}&{$i\neq j\neq k$.}\\\\
&($\vert U_i\rangle, \vert U_j\rangle, \vert {V_i}\rangle, \vert{W_j}\rangle$)
&($\vert U_i\rangle, \vert U_j\rangle, \vert{V_j}\rangle, \vert{W_i}\rangle$)
&($\vert V_i\rangle, \vert V_j\rangle, \vert
{U_i}\rangle,\vert{W_j}\rangle$)\\\\
&($\vert V_i\rangle, \vert V_j\rangle, \vert{U_j}\rangle, \vert{W_i}\rangle$)
&($\vert W_i\rangle, \vert W_j\rangle, \vert{U_i}\rangle, \vert{V_j}\rangle$)
&($\vert{W_i}\rangle,\vert{W_j}\rangle,\vert{U_j}\rangle, \vert{V_i}\rangle$)\\
& {$i,~j\in P$,}&{ $P=\{1,2,3\}$,}&{ $i\neq j$.}\\\\
\hline\\
 {IV$_{3}$}&$(\vert U_i\rangle,\vert U_j\rangle,\vert
{V_i}\rangle,\vert {W_k}\rangle$)& $(\vert U_i\rangle,\vert
U_j\rangle,\vert {V_j}\rangle,\vert {W_k}\rangle$)&
$(\vert U_i\rangle,\vert U_j\rangle,\vert {W_i}\rangle,\vert {V_k}\rangle$)\\\\
&$(\vert U_i\rangle,\vert U_j\rangle,\vert {W_j}\rangle,\vert{V_k}\rangle)$&
$(\vert V_i\rangle,\vert V_j\rangle,\vert{U_i}\rangle,\vert {W_k}\rangle$)&
$(\vert V_i\rangle,\vert V_j\rangle,\vert {U_j}\rangle,\vert
{W_k}\rangle$)\\\\
&$(\vert V_i\rangle,\vert V_j\rangle,\vert {W_i}\rangle,\vert {U_k}\rangle$)
&$(\vert V_i\rangle,\vert V_j\rangle,\vert{W_j}\rangle,\vert {U_k}\rangle$)
&$(\vert W_i\rangle,\vert
W_j\rangle,\vert{U_i}\rangle,\vert{V_k}\rangle$)\\\\
&$(\vert W_i\rangle,\vert W_j\rangle,\vert {U_j}\rangle,\vert {V_k}\rangle$)&
$(\vert W_i\rangle,\vert W_j\rangle,\vert{U_k}\rangle,\vert {V_i}\rangle$)&
$(\vert W_i\rangle,\vert W_j\rangle,\vert {U_k}\rangle,\vert {V_j}\rangle$)\\
&{ $i,~j\in P$,}&{ $P=\{1,2,3\}$,}&{ $i\neq j\neq k$.}\\\\
\hline\\ {IV$_{4}$}&($\vert U_i\rangle,\vert U_j\rangle,\vert
{V_i}\rangle,\vert {V_j}\rangle$) &($\vert U_i\rangle,\vert
U_j\rangle,\vert {W_i}\rangle,\vert {W_j}\rangle$)&
($\vert V_i\rangle,\vert V_j\rangle,\vert{W_i}\rangle, \vert {W_j}\rangle$)\\
& {$i,~j\in P$,}&{ $P=\{1,2,3\}$,}&{ $i\neq j.$}\\\\
\hline\\ {IV$_{5}$}&($\vert U_i\rangle,\vert
U_j\rangle,\vert{V_k}\rangle,\vert {W_k}\rangle$) &($\vert
V_i\rangle,\vert V_j\rangle,\vert {U_k}\rangle,\vert {W_k}\rangle$)
&($\vert W_i\rangle,\vert W_j\rangle,\vert {U_k}\rangle,\vert{V_k}\rangle$)\\
&  {$i,~j,~k\in P$,}&{$P=\{1,2,3\}$,}&{$i\neq j\neq k$.}\\\\
\hline\hline\end{tabular}\end{small}}\end{center}

\noindent  when $a=b=c=d=\frac{1}{\sqrt{6}}$,
$f=\frac{1}{\sqrt{3}}$, and $\varphi=\alpha+\beta-\gamma=2k\pi+\pi$,
of which the corresponding states are equivalent to Type II
entangled states under SLOCC. Furthermore, there is no extremal
value found for the Type V$_{3-6}$ states with all coefficients
nonzero. Therefore, there is no new type of entangled states with
linear combination of five terms; they can all be transformed either
into Type I or Type II states under SLOCC.

For six term cases, there are $(^9_6)=84$ linear combinations, all
of which are genuine entangled states. They are classified into $4$
types as listed in Table 3.

For the Type VI$_{1}$ case, we take
$\vert\Psi\rangle =a \vert U_1\rangle+b e^{i \alpha}\vert
{U_2}\rangle+c e^{i \beta}\vert {U_3}\rangle
+d e^{i \gamma}\vert {V_1}\rangle+
f e^{i \xi}\vert {V_2}\rangle+g e^{i \sigma}\vert {V_3}\rangle$
as an example, where $a,~b,~c,~d,~f,~g\neq 0$ are real and
satisfy the normalization condition
$a^2+b^2+c^2+d^2+f^2+g^2=1$, and
$\alpha,~\beta,~\gamma,~\xi,~\sigma$ are relative phases.
It can be shown that there is an extremal value
with $\eta=0.63093$ when $a=d$, $b=f$,
$c=g$, $\omega=\beta+\gamma-\sigma=2k\pi+\pi$,
$\varphi=\alpha+\gamma-\xi=2k\pi$;
or when $a=g$, $b=f$, $c=d$, $\omega=\beta+\gamma-\sigma=2k\pi$,
$\varphi=\alpha+\gamma-\xi=2k\pi+\pi$, which equals to
the extremal value of the entanglement of Type I
states. Furthermore, it can be proven that
Type VI$_{1}$ states are equivalent to
Type I states under SLOCC.
For the Type VI$_{2}$ case, we take
$\vert\Psi\rangle =a \vert U_1\rangle+b e^{i \alpha}\vert
{U_2}\rangle+c e^{i \beta}\vert {V_1}\rangle
+d e^{i \gamma}\vert {V_3}\rangle+
f e^{i \xi}\vert {W_2}\rangle+g e^{i \sigma}\vert {W_3}\rangle$
as an example, where $a,~b,~c,~d,~f,~g\neq 0$ are real
and satisfy the normalization condition
$a^2+b^2+c^2+d^2+f^2+g^2=1$, and
$\alpha,~\beta,~\gamma,~\xi,~\sigma$
are relative phases.
It can be verified that there is another extremal value
of the entanglement with $\eta=0.78969$
when $a=b=c=d=f=g={1\over{\sqrt{6}}}$, and
$\varphi=\alpha+\beta-\gamma-\xi+\sigma=2k\pi$.
Obviously, the extremal value of $\eta$ in this case
differs from those of Type I and II states.

\begin{center}{\bf Table 2.$~~~$Classification of entangled states
with five terms }\\
\vskip .3cm
\end{center}
{ \begin{tiny}
  \begin{tabular}{cccc}\hline\hline\\
  {\normalsize Types}&&{\normalsize Linear Combinations} \\\\
  \hline\\
{V$_{1}$}&($\vert U_i\rangle,\vert U_j\rangle,\vert {U_k}
\rangle,\vert {V_m}\rangle, \vert {V_n}\rangle$)
&($\vert U_i\rangle, \vert U_j\rangle, \vert {U_k}\rangle,
\vert {W_m}\rangle, \vert {W_n}\rangle$)
&($\vert V_i\rangle, \vert V_j\rangle, \vert {V_k}\rangle,
  \vert {U_m}\rangle, \vert {U_n}\rangle$)\\\\
&($\vert V_i\rangle, \vert V_j\rangle, \vert {V_k}\rangle,
\vert {W_m}\rangle, \vert {W_n}\rangle$)
&($\vert W_i\rangle, \vert W_j\rangle, \vert {W_k}\rangle,
\vert{U_m}\rangle,  \vert {U_n}\rangle$)
&($\vert W_i\rangle, \vert W_j\rangle, \vert {W_k}\rangle,
  \vert {V_m}\rangle, \vert{V_n}\rangle$)\\
&{$i,~j,~k,~m,~n\in P$,}&{$P=\{1,2,3\}$, $i\neq j\neq k$,}&{ $m\neq
n$.}\\\\
&($\vert U_i\rangle, \vert U_j\rangle, \vert {V_i}\rangle,
  \vert{V_j}\rangle,  \vert {W_i}\rangle$)
&($\vert U_i\rangle, \vert U_j\rangle, \vert {V_i}\rangle,
\vert {V_j}\rangle, \vert {W_j}\rangle$)
&($\vert V_i\rangle, \vert V_j\rangle, \vert {W_i}\rangle,
\vert{W_j}\rangle,  \vert {V_i}\rangle$)\\\\
&($\vert V_i\rangle, \vert V_j\rangle, \vert {W_i}\rangle,
\vert {W_j}\rangle, \vert {V_j}\rangle$)
&($\vert V_i\rangle, \vert V_j\rangle, \vert {W_i}\rangle,
\vert {W_j}\rangle, \vert {U_i}\rangle$)
&($\vert V_i\rangle, \vert V_j\rangle, \vert {W_i}\rangle,
\vert {W_j}\rangle, \vert {U_j}\rangle$)\\
&{$i,~j\in P$,}&{$P=\{1,2,3\}$,}& {$i\neq j$,}\\\\
\hline\\ {V$_{2}$}&($\vert U_i\rangle$, $\vert U_j\rangle$, $\vert
{V_i}\rangle$, $\vert {V_j}\rangle$, $\vert {W_k}\rangle$) &($\vert
V_i\rangle$, $\vert V_j\rangle$, $\vert{W_i}\rangle$, $\vert
{W_j}\rangle$, $\vert {V_k}\rangle$) &($\vert V_i\rangle$, $\vert
V_j\rangle$, $\vert {W_i}\rangle$,
  $\vert {W_j}\rangle$, $\vert {U_k}\rangle$)\\
&{$i,~j,~k\in P$,}&{$P=\{1,2,3\}$,}&{$i\neq j\neq k$.}\\\\
\hline\\ {V$_{3}$}&($\vert U_i\rangle$, $\vert U_j\rangle$, $\vert
{U_k}\rangle$, $\vert {V_m}\rangle$, $\vert {W_m}\rangle$) &($\vert
V_i\rangle$, $\vert V_j\rangle$, $\vert {V_k}\rangle$, $\vert
{U_m}\rangle$, $\vert {W_m}\rangle$) & ($\vert W_i\rangle$, $\vert
W_j\rangle$, $\vert {W_k}\rangle$,
$\vert {U_m}\rangle$, $\vert {V_m}\rangle$)\\
&{$i,~j,~k,~m\in P$,}&{$P=\{1,2,3\}$, }&{$i\neq j\neq k$}\\\\
\hline\\ {V$_{4}$}&$(\vert U_i\rangle$, $\vert U_j\rangle$, $\vert
{U_k} \rangle$, $\vert {V_m}\rangle$, $\vert {W_n}\rangle)$ &$(\vert
V_i\rangle$, $\vert V_j\rangle$, $\vert {V_k}\rangle$,
  $\vert {U_m}\rangle$, $\vert {W_n}\rangle)$
&$(\vert W_i\rangle$, $\vert W_j\rangle$, $\vert {W_k}\rangle$,
  $\vert {U_m}\rangle$, $\vert {V_n}\rangle)$\\
&{$i,~j,~k,~m,~n\in P$,}&{$P=\{1,2,3\}$, }&{$i\neq j\neq k$, $m\neq
n$.}\\\\
& $(\vert U_i\rangle$, $\vert U_j\rangle$, $\vert {V_i}\rangle$,
$\vert{V_k}\rangle$, $\vert {W_i}\rangle)$
& $(\vert U_i\rangle$, $\vert U_j\rangle$, $\vert {V_j}\rangle$,
  $\vert {V_k}\rangle$, $\vert {W_j}\rangle)$
& $(\vert U_i\rangle$, $\vert U_j\rangle$, $\vert {W_i}\rangle$,
  $\vert {W_k}\rangle$, $\vert {V_i}\rangle)$\\\\
& $(\vert U_i\rangle$, $\vert U_j\rangle$, $\vert {W_j}\rangle$,
  $\vert {W_k}\rangle$, $\vert {V_j}\rangle)$
& $(\vert V_i\rangle$, $\vert V_j\rangle$, $\vert {W_i}\rangle$,
  $\vert {W_k}\rangle$, $\vert {U_i}\rangle)$
& $(\vert V_i\rangle$, $\vert V_j\rangle$, $\vert {W_j}\rangle$,
$\vert {W_k}\rangle$, $\vert {U_j}\rangle)$\\
&{$i,~j,~k\in P$,}&{$P=\{1,2,3\}$, }&{$i\neq j\neq k$.}\\\\
\hline\\ {V$_{5}$}&($\vert U_i\rangle$, $\vert U_j \rangle$,
$\vert{V_i}\rangle$, $\vert {V_k}\rangle$, $\vert {W_j}\rangle$)
&($\vert U_i\rangle$, $\vert U_j\rangle$, $\vert {V_j}\rangle$,
$\vert {V_k}\rangle$, $\vert {W_i}\rangle$) &($\vert V_i\rangle$,
$\vert V_j\rangle$,
$\vert {W_i}\rangle$, $\vert {W_k}\rangle$, $\vert
{V_j}\rangle$)\\\\
&($\vert V_i\rangle$, $\vert V_j\rangle$,
$\vert {W_j}\rangle$, $\vert {W_k}\rangle$, $\vert {V_i}\rangle$)
&($\vert V_i\rangle$, $\vert V_j\rangle$,
$\vert {W_i}\rangle$, $\vert {W_k}\rangle$, $\vert {U_j}\rangle$)
  &($\vert V_i\rangle$, $\vert V_j\rangle$,
  $\vert {W_j}\rangle$, $\vert {W_k}\rangle$, $\vert {U_i}\rangle$)\\
&{$i,~j,~k\in P$,}&{$P=\{1,2,3\}$, }&{$i\neq j\neq k$.}\\\\
\hline\\ {V$_{6}$}&($\vert U_i\rangle$,$\vert U_j\rangle$,
$\vert{V_i}\rangle$,$\vert {V_k}\rangle$,$\vert {W_k}\rangle$)
&($\vert U_i\rangle$,$\vert U_j\rangle$,$\vert {V_j}\rangle$, $\vert
{V_k}\rangle$,$\vert {W_k}\rangle$) &($\vert V_i\rangle$,$\vert
V_j\rangle$,$\vert {W_i}\rangle$,
$\vert {W_k}\rangle$,$\vert {V_k}\rangle$)\\\\
&($\vert V_i\rangle$,$\vert V_j\rangle$,$\vert {W_j}\rangle$,
$\vert {W_k}\rangle$,$\vert {V_k}\rangle$)
&($\vert V_i\rangle$,$\vert V_j\rangle$,$\vert {W_i}\rangle$,
$\vert {W_k}\rangle$,$\vert {U_k}\rangle$)
&($\vert V_i\rangle$,$\vert V_j\rangle$,$\vert {W_j}\rangle$,
$\vert {W_k}\rangle$,$\vert {U_k}\rangle$)\\
&{$i,~j,~k\in P$,}&{$P=\{1,2,3\}$,}&{$i\neq j\neq k$.}\\\\
\hline\hline\end{tabular}\end{tiny}} \vskip .4cm

\begin{centering}
{\bf Table 3.$~~~$Classification of entangled states
with six terms }\\
\vskip .4cm {\small \begin{tabular}{c} \hline\hline\\
{Types~~~~~~}~~~~~~{~~~~~~~~~~~~~~~~~~~~~~~Linear Combinations~~~~~
~~~~~~~~~~~~~~~~~~~~~~~~~~~}\\\\
 \hline\\
{VI$_{1}$~~~}~~($\vert U_i\rangle,\vert U_j\rangle, \vert
{U_k}\rangle,\vert {V_i}\rangle,\vert {V_j}\rangle,\vert
{V_k}\rangle$)~~~~($\vert U_i\rangle,\vert U_j\rangle,\vert
{U_k}\rangle,
\vert {W_i}\rangle,\vert {W_j}\rangle,\vert {W_k}\rangle$)\\\\
~~~~~~~~~~~~~~($\vert V_i\rangle,\vert V_j\rangle,\vert
{V_k}\rangle, \vert {W_i}\rangle,\vert {W_j}\rangle,\vert
{W_k}\rangle$)~~
{$i,~j,~k\in P$,}~{$P=\{1,2,3\}$, $i\neq j\neq k$.}\\\\
~~ $(\vert U_i\rangle,\vert U_j\rangle,\vert {V_i}\rangle,
\vert{V_j}\rangle,\vert {W_i}\rangle,\vert {W_j}\rangle)$
~~{$i,~j\in P$,}~~{$P=\{1,2,3\}$,~$i\neq j$.} \\\\
\hline\\
{VI$_{2}$~~~}~~($\vert U_i\rangle,\vert U_j\rangle,\vert
{V_i}\rangle, \vert {V_k}\rangle,\vert{W_j}\rangle,\vert
{W_k}\rangle)$ ~~~~($\vert U_i\rangle,\vert U_j\rangle,\vert
{V_j}\rangle,\vert {V_k}\rangle,
\vert{W_i}\rangle,\vert {W_k}\rangle)$\\\\
{$i,~j,~k\in P$,}~~{$P=\{1,2,3\}$,~ $i\neq j\neq k$.~~~~~~
~~~~~~~~~~~~~~~~~~~~~~}\\\\
\hline\\
{VI$_{3}$~~~~~}~~($\vert U_i\rangle,\vert U_j\rangle,\vert
{U_k}\rangle, \vert{V_i}\rangle,\vert {V_j}\rangle,\vert
{W_i}\rangle)$~~~~$(\vert U_i\rangle,\vert U_j\rangle,\vert
{U_k}\rangle,
\vert {V_i}\rangle,\vert {V_j}\rangle,\vert {W_j}\rangle)$~~\\\\
~~~~~~~~~~~$(\vert U_i\rangle,\vert U_j\rangle,\vert {U_k}\rangle,
\vert{W_i}\rangle,\vert {W_j}\rangle,\vert {V_i}\rangle)$~~$(\vert
U_i\rangle,\vert U_j\rangle,\vert {U_k}\rangle,
\vert {W_i}\rangle,\vert {W_j}\rangle,\vert {V_j}\rangle)$\\\\
~~~~~~~~~~~$(\vert V_i\rangle,\vert V_j\rangle,\vert {V_k}\rangle,
\vert{W_i}\rangle,\vert {W_j}\rangle,\vert {U_i}\rangle)$~~~~$(\vert
V_i\rangle,\vert V_j\rangle,\vert {V_k}\rangle,
\vert {W_i}\rangle,\vert {W_j}\rangle,\vert {U_j}\rangle)$\\\\
~~~~~~~~~~$(\vert V_i\rangle,\vert V_j\rangle,\vert {V_k}\rangle,
\vert{U_i}\rangle,\vert {U_j}\rangle,\vert {W_i}\rangle)$~~~~$(\vert
V_i\rangle,\vert V_j\rangle,\vert {V_k}\rangle,
\vert {U_i}\rangle,\vert {U_j}\rangle,\vert {W_j}\rangle)$\\\\
~~~~~~~~~~~~$(\vert W_i\rangle,\vert W_j\rangle,\vert {W_k}\rangle,
\vert{U_i}\rangle,\vert {U_j}\rangle,\vert {V_i}\rangle)$~~$(\vert
W_i\rangle,\vert W_j\rangle,\vert {W_k}\rangle,
\vert {U_i}\rangle,\vert {U_j}\rangle,\vert {V_j}\rangle)$\\\\
~~~~~~~~~~~~$(\vert W_i\rangle,\vert W_j\rangle,\vert {W_k}\rangle,
\vert{V_i}\rangle,\vert {V_j}\rangle,\vert {U_i}\rangle)$~~~$(\vert
W_i\rangle,\vert W_j\rangle,\vert {W_k}\rangle,
\vert{V_i}\rangle,\vert {V_j}\rangle,\vert {U_j}\rangle)$\\\\
{$i,~j,~k\in P$,}~~{$P=\{1,2,3\}$, $i\neq j\neq k$.~~~~~~~
~~~~~~~~~~~~~~~~~~~~~~}\\\\
\hline\\
{VI$_{4}$~~~}~~$(\vert U_i\rangle,\vert U_j\rangle,\vert {U_k}
\rangle,\vert{V_i}\rangle,\vert {V_j}\rangle,\vert {W_k}\rangle)$
~~~~$(\vert U_i\rangle,\vert U_j\rangle,\vert {U_k}\rangle,
\vert {W_i}\rangle,\vert {W_j}\rangle,\vert {V_k}\rangle)$\\\\
~~~~~~~~$(\vert V_i\rangle,\vert V_j\rangle,\vert {V_k}\rangle,
\vert {W_i}\rangle,\vert {W_j}\rangle,\vert {U_k}\rangle)$~~~$(\vert
V_i\rangle,\vert V_j\rangle,\vert {V_k}\rangle,
\vert {U_i}\rangle,\vert {U_j}\rangle,\vert {W_k}\rangle)$\\\\
~~~~~~~~~~~~$(\vert W_i\rangle,\vert W_j\rangle,\vert {W_k}\rangle,
\vert{U_i}\rangle,\vert {U_j}\rangle,\vert {V_k}\rangle)$~~~$(\vert
W_i\rangle,\vert W_j\rangle,\vert {W_k}\rangle,
\vert{V_i}\rangle,\vert {V_j}\rangle,\vert {U_k}\rangle)$\\\\
~~~~{$i,~j,~k\in P$,}~~{$P=\{1,2,3\}$, $i\neq j\neq k$.~~~~~~~
~~~~~~~~~~~~~~~~~~~~~~}\\\\
~~~~~~~~~~~~$(\vert U_i\rangle,\vert U_j\rangle,\vert {V_i}\rangle,
\vert{V_j}\rangle,\vert {W_i}\rangle,\vert {W_k}\rangle)$~~~$(\vert
U_i\rangle,\vert U_j\rangle,\vert {V_i}\rangle,
\vert {V_j}\rangle,\vert {W_j}\rangle,\vert {W_k}\rangle)$\\\\
~~~~~~~~~~~~$(\vert U_i\rangle,\vert U_j\rangle,\vert {W_i}\rangle,
\vert {W_j}\rangle,\vert {V_i}\rangle,\vert {V_k}\rangle)$~~~$(\vert
U_i\rangle,\vert U_j\rangle,\vert {W_i}\rangle,
\vert {W_j}\rangle,\vert {V_j}\rangle,\vert {V_k}\rangle)$\\\\
~~~~~~~~~~~~$(\vert V_i\rangle,\vert V_j\rangle,\vert {W_i}\rangle,
\vert {W_j}\rangle,\vert {U_i}\rangle,\vert {U_k}\rangle)$~~~$(\vert
V_i\rangle,\vert V_j\rangle,\vert {W_i}\rangle,
\vert {W_j}\rangle,\vert {U_j}\rangle,\vert {U_k}\rangle)$\\\\
~~~~{$i,~j,~k\in P$,}~~{$P=\{1,2,3\}$,~$i\neq j\neq k$.~~~~~~~~~
~~~~~~~~~~~~~~~~~~~~}\\\\
\hline\hline\\
\end{tabular}}
\end{centering}

\noindent In fact, it can be proven that Type VI$_{2}$ states
are inequivalent to Type I and II states under
SLOCC. To do so, we take

$$
\vert\Psi\rangle
=\left(x_1\vert11\rangle+x_2\vert10\rangle+x_3\vert01\rangle
+x_4\vert0-1\rangle+x_5\vert-10\rangle\right)+x_6\vert-1-1\rangle
\eqno(5)$$
as an example with nonzero complex coefficients $(x_{1},x_{2},\cdots, x_{6})$,
where the five terms in the parentheses form the Type V$_{5}$
state, which is equivalent to a Type I state, while the
remaining part is always equivalent to a bipartite product state
under SLOCC.  Since all coefficients are nonzero for the case
considered, the state in (5) remains a
linear combination of a Type I entangled state
and a bipartite product state. It will degenerate
into a Type I state only when $x_6\rightarrow0$.
Thus, it is proven that Type VI$_{2}$ is
inequivalent to a Type I state. Furthermore,
(5) can also be written as
$$\vert\Psi\rangle
=\left(\left(x_1\vert11\rangle+x_6\vert-1-1\rangle+x\vert00\rangle\right)\right.$$
$$\left.+\left(x_2\vert10\rangle+x_3\vert01\rangle
+x_4\vert0-1\rangle+x_5\vert-10\rangle\right)-x\vert00\rangle\right),\eqno(6)$$
where $x$ is an arbitrary nonzero complex number.
It is obvious that the terms in the first parentheses form
a Type II state, while the terms in the second parentheses form
a Type V$_{4}$ state. Under SLOCC, a Type V$_{4}$ state is
always equivalent to a Type I state. Therefore, the state in (6)
becomes a linear combination of a Type I and a Type II state
if all coefficients $(x_{1},x_{2},\cdots, x_{6})$ and $x$ are
nonzero, which constitutes a proof that the VI$_{2}$ is
neither equivalent to a Type I state, nor equivalent to
a Type II state under SLOCC. Hence the Type VI$_{2}$ configuration
is  inequivalent to Type I and II states under
SLOCC, which is called Type III.
Similar analyses of Type VI$_{3}$ and VI$_{4}$ configurations was also
carried out, from which we did not find any further new types of entangled
states.

\vskip .2cm \hskip 0.3cm Let $C^3$ denote a Hilbert subspace of
$SU(2)$ spanned by the single qutrit states. Then,
${(C^{3}})^{\otimes N}$ is spanned by the N-particle product states.
In this case, local unitary operations are elements among
$U(1)\otimes{(SU(2))}^{\otimes N}$, where $U(1)$ provides an overall
phase factor. According to the analysis shown in Ref. 16, there are
$2\times 3^{N}$ real parameters needed in a description of any
vector in ${(C^{3}})^{\otimes N}$. Any $SU(2)$ transformation needs
three real parameters to describe it, and there is one more for
$U(1)$. Therefore, the number of independent real parameters needed
in a description of LU inequivalent type of states in the system is
$2\times3^{N}-(3N+1)$. For the bipartite case with $N=2$ considered
in this paper, $11$ real parameters are needed to describe an
arbitrary state. Since $12$ real numbers are needed in order to
describe a state with six terms, our analysis on the classification
of bipartite qutrit states with up to six terms provided is complete
and sufficient. It follows that a similar analysis of states with
more than $6$ terms is not necessary.

\vskip .2cm \hskip 0.3cm
In summary, by using the entanglement measure (1) with
SLOCC transformations, a complete analysis for entangled
bipartite qutrit pure states has been carried out.
Three SLOCC inequivalent types of extremely entangled
bipartite qutrit pure states have been identified by using
constrained maximization. The extremal values of these
three types of entanglement are $\eta=0.63093$ (Type I),
$\eta=1$ (Type II), and $\eta=0.78969$ (Type III), respectively,
corresponding to the following three forms under SLOCC:

$$\vert ~{\rm I}~\rangle ={1\over{\sqrt{2}}}(\vert 11\rangle+\vert
00\rangle),~~
\eta=0.63093,$$

$$\vert~ {\rm II}~\rangle ={1\over{\sqrt{3}}}(\vert 11\rangle+\vert
00\rangle+\vert -1~-1\rangle),~~
\eta=1,$$

$$\vert~ {\rm III}~\rangle ={1\over{\sqrt{6}}}(\vert 11\rangle+\vert
-1~-1\rangle+
\vert 10\rangle+\vert 01\rangle+\vert 0~-1\rangle+\vert -1~0\rangle
),~~
\eta=0.78969.\eqno(7)$$

Our results show that the entanglement measure defined by (1)
is also effective in classifying different genuinely
entangled bipartite qutrit pure states. The most important
evidence is that the number of basic ways
of entanglement equals to the number of extremally
entangled types, which is consistent to the conclusion
shown in Ref. 13. Therefore, extremal entanglement is
a necessary condition in finding different types of entanglement
in multipartite pure state systems under SLOCC.

Support from the US National Science Foundation
(0140300), the Southeastern Universitites Research
Association, the Natural Science Foundation of
China (10175031), the Natural Science Foundation of
Liaoning Province (2001101053), the Education Department
of Liaoning Province (202122024), and the
LSU-LNNU joint research program (C164063) is acknowledged.

\end{document}